\documentclass[%
 aip,
 amsmath,amssymb,
preprint,%
]{revtex4-1}

\usepackage{graphicx}
\usepackage{dcolumn}
\usepackage{bm}

\usepackage[utf8]{inputenc}
\usepackage[T1]{fontenc}
\usepackage{mathptmx}

\begin{document}

\preprint{AIP/123-QED}

\title[]{Finite-size effect of critical penetration of Pearl vortices in narrow superconducting flat rings\\}

\author{N. Kokubo}
\affiliation{Department of Engineering Science, University of
Electro-Communications, Chofu, Tokyo 182-8585, Japan}
 \email{kokubo@uec.ac.jp}
 \author{S. Okayasu}%
\affiliation{Advanced Science Research Center, Japan Atomic Research
Institute, Tokai, Ibaraki 319-1195, Japan\\}
\author{T. Nojima}
\affiliation{Institute for Materials Research, Tohoku University,
Sendai, Miyagi 980-8577, Japan\\}


\date{\today}

\begin{abstract}
We revisit the critical penetration of Pearl vortices in narrow
superconducting flat rings cooled in magnetic fields. Scanning
superconducting quantum interference device microscopy measurements
showed how magnetic field penetrates and vortices are trapped in
flat rings made of amorphous MoGe thin films. Counting the number of
trapped vortices for each image, we found that the vortices are
completely excluded from the ring annulus when the applied field $H$
is below a threshold field $H_{\rm{p}}$: Above this field, the
vortices increase linearly with field. The obtained values of
$H_{\rm{p}}$ depend on the annulus width $w_{\rm{ring}}$ and follow
the relation $\mu_0H_{\rm{p}} = (1.9 \pm 0.1)
\Phi_0/w_{\rm{ring}}^2$ with the superconducting flux quantum
$\Phi_0$. This relationship provides an insight into the effect of
the net-current circulating in the annular region, and also leads to
a precise control to trap or eliminate vortices in flat rings.
\end{abstract}

\maketitle


\section{Introduction\\}
Recent advances in microfabrication techniques provide an
opportunity to manipulate the magnetic flux quantized in units of
$\Phi_0(=h/2e)$ induced in a variety of superconducting micro/nano
structures and devices cooled in magnetic fields. Of particular
interest is the arrangement of quantized vortices (called Pearl
vortices\cite{PearlAPL1964}) in thin films of finite size, which has
attracted considerable academic interest for last two decades. In
addition to the self-energy of the vortex, the vortex-vortex
interaction depends on the size and shape of the
film.\cite{Brandt2005}
The interplay between the intervortex interaction and the
confinement results in unique vortex states, different from the
Abrikosov-vortex lattice in bulk superconductors. These include
vortex polygons,\cite{Buzdin1993} concentric vortex
shells,\cite{PalaciosPRL2000, BaelusPRB2004} and vortex
fusion,\cite{MoshNature1995, BenoistZphys1997, SchweigertPRB1998,
Chibotaru2004} which have been directly observed in imaging
experiments on small superconducting discs,\cite{GrigoPRL2006,
Kokubo2010} squares,\cite{MiskoSUST, Kokubo2014},
triangles,\cite{ZhaoEPL2008, Kokubo2015} pentagons,\cite{HoSUST2013}
and others.\cite{CrenPRL2009, TominagaPRB2013, Dang2018}

In recent years there has been renewed interest in vortices trapped
in small superconductors as one of the control sources for
non-equilibrium excess quasiparticles. While micro/nanostructured
superconductors have been incorporated in various devices for
growing fields such as quantum information processing and metrology,
their performances were (partly) degraded by accumulated excess
quasiparticles. To suppress the overheating in the devices,
vortex-trapped small superconductors can be key elements for tuning
the population of the quasiparticles, leading to the improvement of
quality factor in superconducting
resonators,\cite{NsanzinezaPRL2014} the reduction of the energy
relaxation time of superconducting qubits,\cite{WangNcomm2014} and
the suppression of excessive current in electron
turnstiles.\cite{TaupinNcomms10977} Then, it has become a revisited
question how to control the vortex penetration and subsequent vortex
trap in small superconductors with various shapes.

The critical penetration of vortices in small superconductors has
been discussed through the size dependence of the characteristic
threshold field $H_{\rm{p}}$ for the complete exclusion of vortices
from a thin superconducting strip\cite{Likhalev1972, Maksimova1998,
Stan2004, KuitPRB2008} or a thin superconducting
disc\cite{BenoistZphys1997, Kogan2004, NishioPRL2008} where
electromagnetic properties are governed by Pearl's effective
penetration depth $\Lambda = 2\lambda^2/t (\gg \lambda)$ with
magnetic penetration depth $\lambda$ and film thickness $t (<
\lambda)$.\cite{PearlAPL1964} In the strip the first vortex trap
occurs when magnetic field $\mu_0H$ exceeds $(2\Phi_0/\pi w^2)$
ln$(w/\pi\xi)$ with $w (\ll \Lambda)$ being strip width, which is
set by the energy balance between the self-energy $E_0[\approx
(\Phi_0^2/2\pi\mu_0\Lambda)$ln$(w/\pi\xi)]$ of one vortex trapped in
the middle of the strip and the interaction energy $E_1[\approx
(\Phi_0H/4\Lambda)w^2]$ of the vortex with the screening current
flowing in strip edges.\cite{Maksimova1998} The relationship between
$H_{\rm{p}}$ and $w$ has been studied through vortex imaging
experiments on Nb, NdBa$_2$Cu$_3$O$_y$ and
YBa$_2$Cu$_3$O$_{7-\delta}$ thin strips with a few tenth micron
widths cooled in magnetic fields up to $\sim$ 1 mT.\cite{Stan2004,
KuitPRB2008, SuzukiAPL2000}

The issue of the size dependent $H_{\rm{p}}$ is not trivial when it
comes to a narrow flat ring. In response to applied magnetic field,
the ring has the circulating current $I_1$ (e.g. in counterclockwise
direction near the inner axial edge) induced by fluxoid
$\Phi_{\rm{f}}(=N\Phi_0$ with integer $N$) threading the hole and
the screening current $I_2$ (flowing in clockwise direction near the
outer axial edge) by applied magnetic field. The superposition of
the two currents results in the counter current flow in the annular
region between inner $a$ and outer radii $b (> a)$. This resembles
the situation for the strip which carries equal and opposite
currents in edges. Therefore, one might expect naively that the
threshold field for the complete vortex exclusion from the ring
annulus depends on the annulus width
$w_{\rm{ring}}(=b-a)$.\cite{Kogan2004, Brandt2004} However, $I_1$
varies non-monotonously with $H$ due to transitions from the fluxoid
state $N$ to $N\pm1$, while the field dependence of $I_2$ is
monotonous. This gives a stark contrast to the narrow strip,
implying that the effect of the \emph{net} current
$I_{\rm{net}}(=I_1-I_2)$ should be taken into account to determine
$H_{\rm{p}}$. The situation is also different from a slitted
ring/loop, where the net current is interrupted by a slit and the
condition $I_{\rm{net}}=0$ holds.\cite{Brandt2004, Brandt2005}

So far much experimental effort has been devoted to the issue of the
vortex (flux) trap in slitted superconducting loops for improving
the performance of superconducting quantum interference devices
(SQUIDs),\cite{SuzukiAPL2000, Wordenweber2002} while experiments on
a simple narrow ring/loop (with \emph{no} slit) have been
limited.\cite{MillPRB2015} In this study, we report the direct
observation of Pearl vortices trapped in narrow flat rings of
amorphous superconducting films with different sizes by scanning
SQUID microscopy (SSM). Different from the previous study made on
square loops,\cite{Mitsuishi2018} the present data on the flat rings
are able to examine the relationship between $H_{\rm{p}}$ and
$w_{\rm{ring}}$ without ambiguities arisen from the nonuniform width
in loops. Our quantitative analysis shows that a fluxoid state with
negligibly small net current gives a significant contribution to the
first vortex trap.

\section{Experimental}

We prepared amorphous Mo$_x$Ge$_{1-x}$ (MoGe) thin films with $x
\approx$ 80 $\pm$ 2 $\%$, which were sputtered in argon gas
atmosphere on water-cooled Si (100) substrates from the target
composed of high purity germanium pieces (99.999$\%$) glued on top
of a high purity molybdenum (99.99$\%$) plate. The uniformity in
molybdenum (or germanium) distribution was confirmed by electron
probe microanalysis with JOEL JXA-8530F. The film thickness $t$ =
0.21 $\mu$m was determined by measuring the vertical profile of the
sample edge by a stylus surface profiler. Using ultraviolet
lithographic and chemical etching techniques, the films were partly
patterned into Hall bars to determine the superconducting transition
temperature $T_{\rm{c}} (\approx 7.4$ K), the normal resistivity
($\approx$ 1.4 $\mu\Omega$m at 10 K), and the second critical field
$H_{\rm{c2}}$. Then, we estimated the zero-temperature magnetic
penetration depth $\lambda(0) \approx$ 0.46 $\mu$m and the
zero-temperature coherence length $\xi(0) \approx$ 4.4 nm from dirty
limit expressions.\cite{Kes1983} Because the films are thinner than
the penetration depth ($t <\lambda$), their electromagnetic
properties are governed by $\Lambda$ rather than $\lambda$. The rest
of the films were partly patterned into flat rings, as shown in
Fig.1a, with different combinations of inner $a$ and outer radii $b$
for SSM imaging experiments. To reduce possible damage during the
scanning, we deposited 0.1 $\mu$m thick silicon-oxide film on top of
the rings. The parameters of five samples we focus in present study
are summarized in Table 1. For all the samples, the values of $a$
are fixed with $\approx$ 10 $\mu$m, while those of $b$ are changed.
This enables us to examine the effect of the annulus width
$w_{\rm{ring}}$ on the critical penetration of Pearl vortices in
flat rings.

\begin{table}[htb]
\caption{Parameters of amorphous MoGe superconducting thin rings}
  \begin{tabular}{ccccc} \hline \hline
      & $a(\mu$m) & $b(\mu$m)  & $w_{\rm{ring}}(\mu$m) & $\mu_0H_{\rm{p}}(\mu$T) \\ \hline
    C1 & $10.0 \pm 0.5$ & $19.5 \pm 0.5$ & $9.5 \pm 0.7$ & $37 \pm 1 $\\
    C2 & $10.5 \pm 0.5$ & $18.5 \pm 0.5$ & $8.0 \pm 0.7$ & $62 \pm 2 $ \\
    C3 & $10.0 \pm 0.5$ & $24.5 \pm 0.5$ & $14.5 \pm 0.7$ & $20.2 \pm 0.4$ \\
    C4 & $10.0 \pm 0.5$ & $34.5 \pm 0.5$ & $24.5 \pm 0.7$ & $7.8 \pm 0.1$ \\
    C5 & $10.5 \pm 0.5$ & $38.5 \pm 0.5$ & $28.0 \pm 0.7$ & $5.4 \pm 0.2$\\ \hline \hline
  \end{tabular}
\end{table}

We used a scanning SQUID microscope (SQM-2000, SII Nanotechnology)
with a sensor chip integrating a superconducting pickup coil with
the effective diameter of $\approx 9$ $\mu$m and niobium-based
Josephson junctions. The sensor chip was mounted on a
phosphor-bronze cantilever and tilted slightly with respect to the
sample stage. By manipulating motorized $xyz$ precision positioning
devices assembled under the sample stage, the sample surface was
softly in contact with a corner of the sensor chip and scanned in
$x(y)$ direction during the image acquisition. Due to the weak
pinning properties of amorphous MoGe films, the distance between the
pickup coil and the sample surface is an important parameter to be
controlled as the movement of the pick-up coil can drag and/or kick
out vortices during the scanning.\cite{Kokubo2010} To reduce the
coupled motion of vortices, we kept the distance of $\sim$ 5 $\mu$m
between the pickup coil and the sample surface. This allowed us to
image individual vortices in amorphous thin films with reasonable
lateral resolutions when the vortex density is low (see the
Appendix). Small thermal drift present in the sample stage may
distort SSM images and this occurs likely if the large area was
scanned with a step size of 1 $\mu$m. To exclude this drawback in
our setup, all the SSM images on rings in the present study were
taken with a 4 $\mu$m step size. The sample stage has a multi-turn
wound coil for applying small magnetic field $H$ perpendicular to
the sample surface. The whole assembly including the sensor chip and
the sample stage was covered with a $\mu$-metal shield. The ambient
magnetic field around the sample space was reduced to $\approx$ 1
$\mu$T, which was subtracted from the magnitude of applied magnetic
fields.

\section{Results and Discussion}

Figures 1b-1i show a set of SSM images of the ring sample C4
($w_{\rm{ring}}=$ 24.5 $\mu$m) after cooling to 4.0 $\pm$ 0.1 K in
applied magnetic fields from 2 to 9 $\mu$T with 1 $\mu$T field step.
A color bar indicates the magnitude of the magnetic flux
$\Phi_{\rm{s}}$ through the pickup coil. The flux expulsion observed
as a ring allows us to find the position of the sample in each
image. One can also find that a dome like magnetic profile appears
around the hole center marked with a cross, the intensity of which
is larger than that outside the ring. This is known as a result of
field focusing into the hole that occurs when small magnetic field
(below $H_{\rm{p}}$) is applied perpendicularly to a flat ring with
finite screening.\cite{Brandt2004, Kirtly2003} There occurs magnetic
field penetrations from the inner and outer axial edges (which lead
to the counter circulating flow in the annular region), and no
isolated magnetic flux is observed in the ring annulus up to 7
$\mu$T. Meanwhile, as shown in Fig. 1h, one can recognize clearly
the sudden appearance of a magnetic flux spot in lower right of the
annulus at 8 $\mu$T. Then, another flux spot emerges in the opposite
side of the annulus (see Fig. 1i) at a slightly higher field of 9
$\mu$T. Subsequently, the number of flux spots increased one by one
with magnetic field. Because each spot lies within the annular
region, it can be naturally regarded as a Pearl vortex.

\begin{figure}[t]
\begin{center}
\includegraphics[width=20pc]{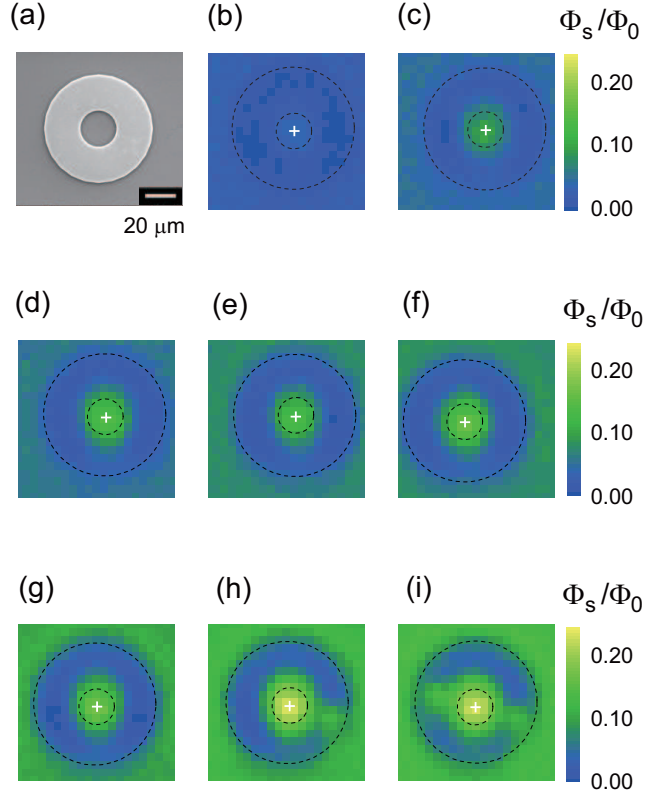}\hspace{2pc}
\caption{Scanning SQUID microscopy images of C4 ring after cooling
to 4.0$\pm$ 0.1 K in different magnetic fields of (b) 2, (c) 3, (d)
4, (e) 5, (f) 6, (g) 7, (h) 8, (i) 9 $\mu$T, respectively. All
images are the same in size of 84 $\times$ 84 $\mu$m$^2$. For each
image hole center is marked with cross.  $\Phi_{\rm{s}}$ is magnetic
flux through pickup coil. Electron micrograph of C4 ring is given in
(a).}
\end{center}
\end{figure}

In order to clarify the magnetic flux trapped in the ring annulus,
we take the difference between images. An example is shown in Fig. 2
where the part of the image data with the flux spot at 9 $\mu$T
(Fig. 1i) has been subtracted from that at 8 $\mu$T (Fig. 1h) with
the flux-free annular region.\cite{subtraction} One can see that the
magnetic flux trapped in the annulus is clearly visible, while the
flux focused in the hole is largely reduced. We plot the profile of
magnetic flux intensity $\Phi_{\rm{s}}$ along the cross section near
the flux center, represented by a solid line (which corresponds to
the scan direction of the pickup coil). It shows a broad peak in the
annular region. HWHM (Half-width of the half maximum) of the peak is
$\approx$ 7 $\mu$m which is wider than the effective penetration
depth $\Lambda$(4 K)[=$\Lambda(0)/(1-(T/T_{\rm{c}})^4)] \approx$ 2.3
$\mu$m.\cite{NishioKPRB2008} This originates from the combined
effect of the spread of the magnetic flux at the measurement
position apart form the sample surface and the finite size
($\approx$ 9 $\mu$m) of the pickup coil. The profile of
$\Phi_{\rm{s}}(r)$ in Fig. 2 can be qualitatively reproduced by a
monopole model because the condition $(r^2+z^2) \gg \Lambda^2$,
where $r(=\sqrt{x^2+y^2})$ and $z$ respectively are the in-plain
distance and the height of the pickup coil from the flux center, is
fulfilled.\cite{WynnPRL2001, NishioMPRB2008, GePRB2013}
The magnetic field $B_{\rm{z}}(r,z)$ (perpendicular to the sample
surface) originating from the magnetic monopole $\Phi_{\rm{m}}$
can be expressed as
\begin{eqnarray*}
B_{\rm{z}}(r,z)=\frac{\Phi_{\rm{m}}}{2\pi}\frac{z+\Lambda}{r^2+(z+\Lambda)^2}.
\end{eqnarray*}
As pointed out by Wynn \emph{et al.},\cite{WynnPRL2001} this model
remains still acceptable as a good approximation even at $r$=0,
provided that
$z$ is larger than the effective penetration depth, i.e., $z
> \Lambda$. Integrating $B_{\rm{z}}(r,z)$ over the effective area $S$
of the pickup coil (assuming a 9 $\mu$m diameter perfect circle as
the pickup coil), we obtain the total flux $\Phi_{\rm{s}}$ through
the coil as function of $r$. Under the condition of
$\Phi_{\rm{m}}=\Phi_0$, the best fit to the flux profile (red line
in Fig. 2) gives $z+\Lambda=$ 7.8 $\pm$ 0.2, resulting in $z= 5.5
\pm$ 0.2 $\mu$m, which is very close to the actual height of the
pickup coil. We note that both width and magnitude of the flux
profile are quantitatively close to our previous result of
individual Pearl vortices in amorphous MoGe thin
films.\cite{NishioKPRB2008} Therefore, each flux spot trapped in the
ring annulus corresponds to a Pearl vortex accompanying the flux
quantum $\Phi_0$.

\begin{figure}[t]
\begin{center}
\includegraphics[width=15pc]{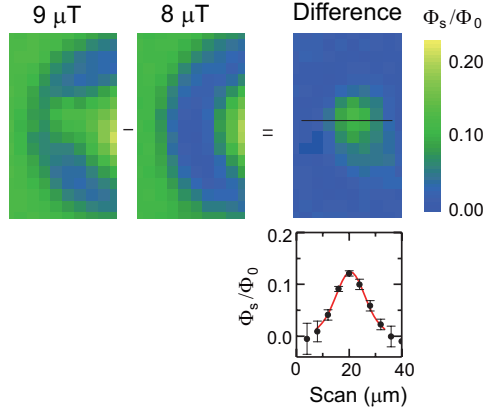}\hspace{2pc}
\caption{Difference between two SSM images of 8 $\mu$T (Fig. 1h) and
9 $\mu$T (Fig. 1i). Profile of flux intensity along the cross
section is also given. }
\end{center}
\end{figure}

To determine the threshold field $H_{\rm{p}}$ for the critical
penetration of the C4 ring, we count the number $N_{\rm{V}}$ of
trapped flux (vortices) in the ring annulus in each image and plot
it against applied field in Fig. 3. One can see that $N_{\rm{V}}$
for the C4 ring (black symbols) increases linearly with applied
field. This behavior can be approximated as
$N_{\rm{V}}=\mu_0(H-H_{\rm{p}})A/\Phi_0$ with the area
$A=\pi(b^2-a^2) \approx$ 3400 $\mu $m$^2$ of the annulus. Then, the
threshold field $\mu_0H_{\rm{p}} \approx$ 7.8 $\mu$T is determined
by the linear extrapolation to $N_{\rm{V}}=0$, as in a previous
study made on Nb strips.\cite{Stan2004} As well as the C4 ring, the
linear approximation fits nicely the data obtained in the C5 (C3)
ring with wider (narrower) width and results in the lower (higher)
threshold field $\mu_0H_{\rm{p}} \approx$ 5.4 (20.2) $\mu$T (see
Fig. 3). Thus obtained values of $H_{\rm{p}}$ are listed in Table 1.

\begin{figure}[t]
\begin{center}
\includegraphics[width=15pc]{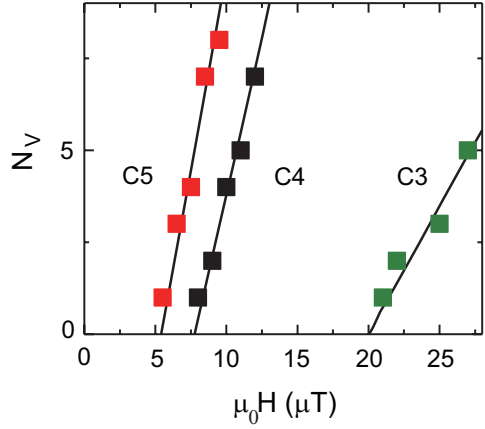}\hspace{2pc}
\caption{Magnetic field vs. number $N_{\rm{V}}$ of trapped flux
(vortices) for three rings }
\end{center}
\end{figure}

We plot the experimentally obtained values of $H_{\rm{p}}$ against
$w_{\rm{ring}}$ in Fig. 4. One can see from this log-log plot that
the data follow the relationship of the form $\mu_0H_{\rm{p}} =
C_{\rm{ring}} \Phi_0/w_{\rm{ring}}^2$, being in good agreement with
the naive expectation.\cite{Brandt2004} The prefactor was obtained
as $C_{\rm{ring}}= 1.9 \pm 0.1 $ from the least square fit of the
data to the relationship as represented by a solid line. Thus
obtained relationship does hold even when using different
definitions for $H_{\rm{p}}$ (, e.g. the lowest field for one Pearl
vortex trapped in the ring annulus), although the prefactor
$C_{\rm{ring}}$ changes slightly.

\begin{figure}[t]
\begin{center}
\includegraphics[width=15pc]{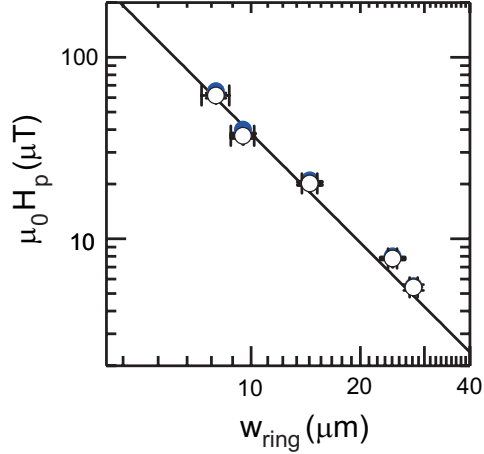}\hspace{2pc}
\caption{Plot of threshold field $H_{\rm{p}}$ vs ring width
$w_{\rm{ring}}$. Open symbols indicate data determined by linear
extrapolation to $N_{\rm{V}}=0$ (see Fig. 3), while solid ones by
the lowest field for one vortex trapped in ring annulus.}
\end{center}
\end{figure}

Let us discuss the prefactor $C_{\rm{ring}}$ of the relationship by
taking into account the net current $I_{\rm{net}}$ in the annular
region. The first vortex trap in the ring annulus can be sensitively
affected by the net force $\Phi_0I_{\rm{net}}$ exerted on the
vortex. Therefore, it is reasonable to consider the situation for
the \emph{zero}-net current in the ring and incorporate it into the
critical penetration. This corresponds to the circularly symmetric
situation of $I_1=I_2$ with the fluxoid state $N$ and may occur when
$H_{\rm{p}}=NH_0$ with a characteristic field $\mu_0H_0\equiv
\Phi_0/A_{\rm{eff}}$, which is set by the effective area
$A_{\rm{eff}} (> \pi a^2)$ of the hole.\cite{BrojenyPRB2003,
Brandt2004} Using the above relationship, we obtain
\begin{eqnarray*}
C_{\rm{ring}}=Nw_{\rm{ring}}^2/A_{\rm{eff}}.
\end{eqnarray*}
Thus, the prefactor is given by the fluxoid number $N$ at
$H_{\rm{p}}$ multiplied with the ratio of squared annulus width and
the effective area. In order to test this conjecture, we
quantitatively estimate the fluxoid number $N$ and its effective
area $A_{\rm{eff}}$ by following the analysis by Brandt and
Clem.\cite{Brandt2004} They numerically calculated field and current
profiles of a flat ring for arbitrary values of
$\Lambda'(T)(=\Lambda'(T)/2)$ and $a/b$, covering the shape of our
rings studied. For the C4 ring, we find the fluxoid number
$N=\lfloor H_{\rm{p}}/H_0 +0.5 \rfloor = $ 4 with the characteristic
field $\mu_0 H_0 =\Phi_0/
A_{\rm{eff}}=\Phi_0\alpha_{\rm{I}}/ab\beta_{\rm{I}} \approx$ 1.8
$\mu$T and the effective area
$A_{\rm{eff}}=ab\beta_{\rm{I}}/\alpha_{\rm{I}} \approx
1.1\times10^{-9}$ m$^2$. Here, we use dimensionless factors of
$\alpha_{\rm{I}} \approx 0.32$ and $\beta_{\rm{I}} \approx 1.05$
 determined respectively from numerical results given in Figs. 6 and
12 of Ref. 30
, provided that $\Lambda'(T)/b \approx 0.032$ and $a/b \approx$
0.29. Substituting these to the above relation, we find that
$Nw_{\rm{ring}}^2/A_{\rm{eff}} \approx 2.1$ which is close to the
experimentally obtained value of $C_{\rm{ring}}=1.9 \pm 0.1$. It
turns out that the products $Nw_{\rm{ring}}^2/A_{\rm{eff}}$ for the
other rings agree  with the experimental value $C_{\rm{ring}}$
within error bars. These reasonable coincidences support the
aforementioned conjecture that the critical penetration of Pearl
vortices in the ring annulus needs the zero-net-current condition.

We comment on the temperature dependence of the fluxoid number $N$
threading the hole. Near $T_c$ the effective penetration depth
becomes divergently large ($\Lambda'/b \gg 1$). Thus, one can use
the dirty limit expression for the effective area $A_{\rm{eff}}
=\pi(b^2-a^2)/2$ln$(b/a)$.\cite{Brandt2004} Using parameters for the
C4 ring, we find $N(H_{\rm{p}}, T_c)=\lfloor H_{\rm{p}}/H_0(T_c)
+0.5 \rfloor = $ 5 which is larger than that at $T =$ 4 K estimated
above. This may have an important consequence for the critical
penetration in the field-cooled ring, as the excess flux should be
excluded from the hole during cooling from $T_c$ to 4 K and can be
subsequently trapped as one vortex
in the ring annulus at lower temperatures (not close to $T_c$).
Thus, the temperature dependence of $N$ can lead to the reasonable
process of the vortex trap in the flat ring cooled at the threshold
field.

Finally, we discuss the critical penetration of Pearl vortices in
narrow square loops. The previous SSM imaging experiments on narrow
square loops made of amorphous MoGe and Nb thin films have reported
the size dependent threshold field.\cite{Mitsuishi2018} The obtained
values of $H_{\rm{p}}$ obey $\mu_0H_{\rm{p}}= C \Phi_0/w^2$, where
the prefactor $C$ is 3.9 when one takes the widest spacing $w_1(=w)$
along the diagonal of the square or 1.8 when the narrowest one
$w_2(=w)$. These prefactors are also explainable when considering
the situation of the zero-net current in loops. Following the
aforementioned analysis, together with the numerical results of
field and current profiles of a square loop,\cite{Brandt2005} we
find that $Nw_1^2/A_{\rm{eff}} \approx$ 3.7 and $Nw_2^2/A_{\rm{eff}}
\approx$ 1.4 for the widest and the narrowest spacings of the MoGe-A
loop, respectively.\cite{parameters} These products, together with
ones obtained in other loops, turn out to be close to the prefactors
in the above relations. Therefore, we believe that narrow flat rings
and square loops share the same mechanism of the first vortex trap.

\section{Conclusion}

To summarize, we have presented SSM images of Pearl vortices trapped
in narrow flat rings made of amorphous MoGe superconducting thin
films cooled in different magnetic fields. Our data showed clearly
the presence of a threshold field $H_{\rm{p}}$, above which the
vortices are trapped in the ring annulus and increase linearly with
applied field. The experimentally obtained values of $H_{\rm{p}}$
depend on the annulus width $w_{\rm{ring}}$ and obey the
relationship of the form $\mu_0H_{\rm{p}} = C_{\rm{circle}}
\Phi_0/w_{\rm{ring}}^2$ with $C_{\rm{circle}}=1.9 \pm 0.1$.
Quantitative analysis on the prefactor $C_{\rm{circle}}$ revealed
that the critical penetration in a flat ring occurs when the net
force exerted on the vortex in the ring annulus becomes negligibly
small. These findings are useful for trapping or eliminating Pearl
vortices in flat rings, which can be crucial elements for designing
various devices for quantum information processing, memory and
metrology.

\begin{acknowledgments}
N. K acknowledges M. Mitsuishi for technical assistance. This work
was supported by JSPS KAKENHI (Grant Numbers 26287075 and 17K05537),
and the Inter-university Cooperative Research Program of the
Institute for Materials Research, Tohoku University (Proposal No.
17K0051 and 18K0012).
\end{acknowledgments}

\appendix*
\section{}

The rest of the amorphous MoGe films used in the present study was
patterned into other shapes including disks with different
diameters. Figure 5a shows a SSM image on a 90 $\mu$m diameter disk
cooled in 8.2 $\mu$T. The image was taken with a 1 $\mu$m step size.
One can see magnetic flux spots with well spaced (no strong
overlaps) and nearly equal magnitude in the disk. They form the
triple shell configuration characterized by (1,6,11) which
represents that the most inner shell has one (Pearl) vortex, the
middle shell is formed by 6 vortices and the most outer shell by 11
vortices.\cite{Kokubo2010} Figure 5b shows an image on the same disk
cooled slightly higher field of 8.4 $\mu$T taken with a 4 $\mu$m
step size. Despite large pixel blocks, one can reasonably find how
vortices are arranged, supporting that each flux spot observed in
Figs. 1h and 1i represents a Pearl vortex even with 4 $\mu$m
step-size measurements. The evolution of quasi-symmetric concentric
vortex shells as function of vorticity was obtained from the set of
SSM images on the same disk cooled in different magnetic
fields.\cite{Kokubo2010} This excludes unintended inhomogeneity like
cracks in the films used in the present study.

\begin{figure}[t]
\begin{center}
\includegraphics[width=20pc]{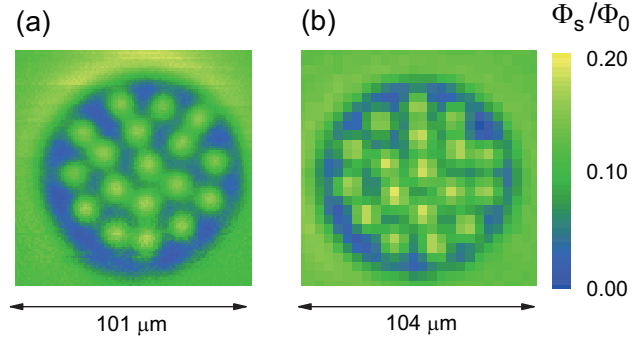}\hspace{2pc}
\caption{Scanning SQUID microscopy images of 90 $\mu$m diameter disk
after cooling to 3.2$\pm$ 0.1 K in magnetic fields of 8.2 $\mu$T (a)
and 8.4 $\mu$T (b), which were taken with different step sizes of 1
$\mu$m and 4 $\mu$m, respectively. }
\end{center}
\end{figure}


\begin{thebibliography}{00}
\bibitem{subtraction} For the subtraction a careful adjustment was made with respect to
sample position in the scale of lateral step size of 4 $\mu$m.
\end{thebibliography}

\end{document}